\documentclass{emulateapj}
\usepackage{apjfonts}

\newcommand\HI{H\,{\sc i}}

\def\kms{km~s$^{-1}$}
\def\etal{{\rm et~al.\ }}

\def\psr{B1757--24}
\def\pwn{G5.27--0.90}
\def\snr{G5.4--1.2}

\shorttitle{PROPER MOTION FOR PSR~B1757--24 NEAR SNR~G5.4--1.2}
\shortauthors{BLAZEK ET AL.}

\begin{document}
\title{The Duck Redux: An Improved Proper Motion Upper Limit
for the \\ Pulsar B1757--24 Near the Supernova Remnant G5.4--1.2}
\submitted{Accepted to {\em The Astrophysical Journal}}
\author{J. A. Blazek,\altaffilmark{1} B. M. Gaensler,\altaffilmark{1,2,3} 
S.  Chatterjee,\altaffilmark{4,1} E. van der Swaluw,\altaffilmark{5} 
F.~Camilo\altaffilmark{6} and B.~W.~Stappers\altaffilmark{7,8}}

\altaffiltext{1}{Harvard-Smithsonian
Center for Astrophysics, 60 Garden Street, Cambridge, MA 02138;
blazek@post.harvard.edu}
\altaffiltext{2}{Alfred P.\ Sloan Research Fellow}
\altaffiltext{3}{Present address: School of Physics A29, The University
of Sydney, NSW 2006, Australia; bgaensler@usyd.edu.au}
\altaffiltext{4}{Jansky Fellow, National Radio Astronomy Observatory, 520 Edgemont Road,
Charlottesville, VA 22903}
\altaffiltext{5}{Royal Netherlands Meteorological Institute (KNMI),
PO Box 201, 3730 AE De Bilt, The Netherlands}
\altaffiltext{6}{Columbia Astrophysics Laboratory, Columbia University,
550 West 120th Street, New York, NY 10027}
\altaffiltext{7}{Stichting ASTRON, Postbus 2, 7990 AA Dwingeloo, The 
Netherlands}
\altaffiltext{8}{Astronomical Institute ``Anton Pannekoek'', University
of Amsterdam, Kruislaan 403, 1098 SJ Amsterdam, The Netherlands}

\begin{abstract}

``The Duck'' is a complicated non-thermal radio system, consisting
of the energetic radio pulsar \psr, its surrounding pulsar wind
nebula \pwn\ and the adjacent supernova remnant (SNR) \snr.  PSR~\psr\
was originally claimed to be a young ($\approx15\,000$~yr) and
extreme velocity  ($\ga1500$~\kms) pulsar which had penetrated
and emerged from
the shell of the associated
SNR~\snr, but recent upper limits on the pulsar's motion have
raised serious difficulties with this interpretation.  We here
present 8.5~GHz interferometric observations of the nebula \pwn\ over a
12-year baseline, doubling the time-span of previous measurements.
These data correspondingly allow us to halve the previous
upper limit on the nebula's westward motion to 14~milliarcseconds yr$^{-1}$
(5-$\sigma$), allowing a substantive reevaluation of this puzzling
object.  We rule out the possibility that the pulsar
and SNR were formed from a common supernova explosion
$\approx15\,000$~yrs ago as implied by the pulsar's characteristic age,
but conclude that an old ($\ga70\,000$~yr) pulsar
/ SNR association, or a situation in which the pulsar and SNR are
physically unrelated, are both still viable explanations.

\end{abstract}


\section{Introduction}

The formation and subsequent evolution of pulsars are
not yet fully understood. A powerful constraint
on these processes is provided by an independent age estimate.
In cases where both a pulsar and its associated
supernova remnant (SNR) can be identified, an age
estimate which is independent of both distance and inclination
effects
is simply $t_p = \Theta / \mu$,
where $\Theta$ is the angular separation between the pulsar
and its inferred birth site, and $\mu$ is the pulsar's proper motion
\citep{mgb+02,klh+03}. 
This can then be compared to the age of the system
as expected from spin-down \citep[e.g.,][]{lk05}:
\begin{equation}
t_p=\frac{P}{(n-1)\dot{P}}\left[1-\left(\frac{P_0}{P}\right)^{n-1}\right] ,
\label{eqn_age}
\end{equation}
where $P$ is the current spin period, $\dot{P}$ is the 
time-derivative of $P$, $P_0$ is the period at birth,
and $n$ is the ``braking index'' (see further
discussion in \S\ref{sec_old}). Comparison of
these two independent age estimates  provides
information on $P_0$ and $n$, i.e., on the processes which
impart and then dissipate the considerable angular momentum of neutron stars
 (e.g., Gaensler \& Frail 2000,
hereafter GF00; Kaspi \etal\ 2001b\nocite{gf00,krv+01}).
If one assumes that 
$P_0\ll P$ and $n=3$, Equation~(\ref{eqn_age}) reduces to
the expression for the ``characteristic age'' of a pulsar, $\tau_c \equiv 
P/2\dot{P}$.

PSR~\psr\ is an isolated 125-ms pulsar, surrounded by the cometary
radio and X-ray pulsar wind nebula (PWN) \pwn, which in turn is
located just outside the SNR~\snr\
\citep{ckk+87,mkj+91,fk91,fkw94,kggl01}.
As shown
in Figure~\ref{fig_snr}, 
the combined system has a distinctive morphology which has led to
it being termed ``the Duck''. Because of the proximity of the pulsar
to the SNR, and 
because the PWN morphology suggests that the pulsar is
moving away from the SNR interior, it has been widely assumed that
the pulsar and the SNR are physically associated.  The SNR and PWN
also have very similar \HI\ absorption spectra, both suggesting a 
distance of $\sim5$~kpc \citep{fkw94}, and consistent with the distance
of 5.1~kpc implied by the pulsar's dispersion measure and the
Galactic electron density model of \cite{cl02}.  We consequently
adopt a common distance of 5~kpc for pulsar, PWN and SNR in further
discussion.

PSR~\psr\ has a characteristic age $\tau_c = 15$~kyr \citep{mhth05}.
If we assume $t_p = \tau_c$ and that the pulsar has traveled $\Theta
= 16\farcm1 -20\farcm6$ from the SNR's center in its lifetime
\citep{fkw94},\footnote{The range of values quoted for $\Theta$
reflects the uncertainty in locating the pulsar's presumed birthplace.
This results
from the fact that the pulsar's inferred trajectory does
not pass through the SNR's geometric center 
\citep[Fig.~\ref{fig_snr};][]{fkw94}. See
\S\ref{sec_old} for further discussion.}
this implies a westward proper motion for the pulsar of magnitude
$\mu=\Theta/\tau_c = 63-80$~mas~yr$^{-1}$.  However, radio
interferometric observations of the western tip of the PWN taken
at the Very Large Array (VLA) over 6.7~years yielded a surprising
5-$\sigma$ upper limit $\mu< 25$~mas~yr$^{-1}$ (GF00\nocite{gf00}),
implying a pulsar age $t_p > 39-50$~kyr~$\gg\tau_c$, if the pulsar
was born near the SNR's geometric center.  GF00 used the stand-off
distance of the PWN, the radius of the SNR and the separation of
the pulsar from the SNR's center to derive a solution for the
system's evolution which predicted $t_p \sim 90- 170$~kyr and $\mu
\sim 10$~mas~yr$^{-1}$.  Subsequently, Thorsett \etal\ (2002,
hereafter TBG02\nocite{tbg02}) observed the pulsar itself with the
VLA over a 3.9-year baseline, and placed an independent upper limit
on westward proper motion\footnote{TBG02 state that their limit is
$\mu < 16$~mas~yr$^{-1}$ at 95\% confidence (i.e., $2\sigma$).  Here
we use 5-$\sigma$ limits throughout, and have adjusted their limit
accordingly.} of $\mu < 37$~mas~yr$^{-1}$.  TBG02 argue that this
is most easily explained if PSR~\psr\ and SNR~\snr\ are unrelated,
and if the pulsar is instead moving away from the center of the
PWN~\pwn. Assuming $t_p \approx \tau_c$, one can then predict a
proper motion $\mu \sim 5$~mas~yr$^{-1}$.  As a further alternative,
\cite{gva04b} has argued that this system results from a massive
high-velocity progenitor star which went supernova inside its
wind-blown bubble. 
The pulsar began its life substantially offset
from the cavity's center, but 
the resulting SNR expands to take on the shape
of the cavity.  \cite{gva04b} subsequently develops a
model in which $t_p \approx 54$~kyr, $\Theta \approx 6'$ and $\mu
\approx 7$~mas~yr$^{-1}$.

As an attempt to resolve this puzzling situation, we have conducted a
new observation of the western cometary tip of PWN~\pwn\ (see inset to
Fig.~\ref{fig_snr}), which doubles the time baseline considered by GF00
to 12 years.  In \S\ref{sec_obs} we present our new observations and
corresponding measurement of proper motion, while in \S\ref{sec_disc} we
we interpret these new results in the context of various possibilities
proposed for the origin and evolution of this system.

\section{Observations and Analysis}
\label{sec_obs}

We have utilized three separate  observations of \pwn, all
carried out near 8.5~GHz with the VLA in its hybrid BnA configuration.
The first two epochs were on 1993~Feb~02 and 1999~Oct~23, as discussed
by GF00; the third epoch was on 2005~Jan~22, with an on-source
integration time of 2.9~hr, and parameters otherwise the same as
for the 1999 epoch reported by GF00.  In particular, observations
of \pwn\ at all three epochs were phase-referenced to the source
PMN~J1751--2524, located $2\fdg1$ from \pwn\ (and for consistency
using the same position for  PMN~J1751--2524 at each epoch, even
though the best estimate of the position of this source has been
updated by a small amount in recent years).

\begin{figure}
\centerline{\includegraphics[width=0.9\textwidth,angle=270]{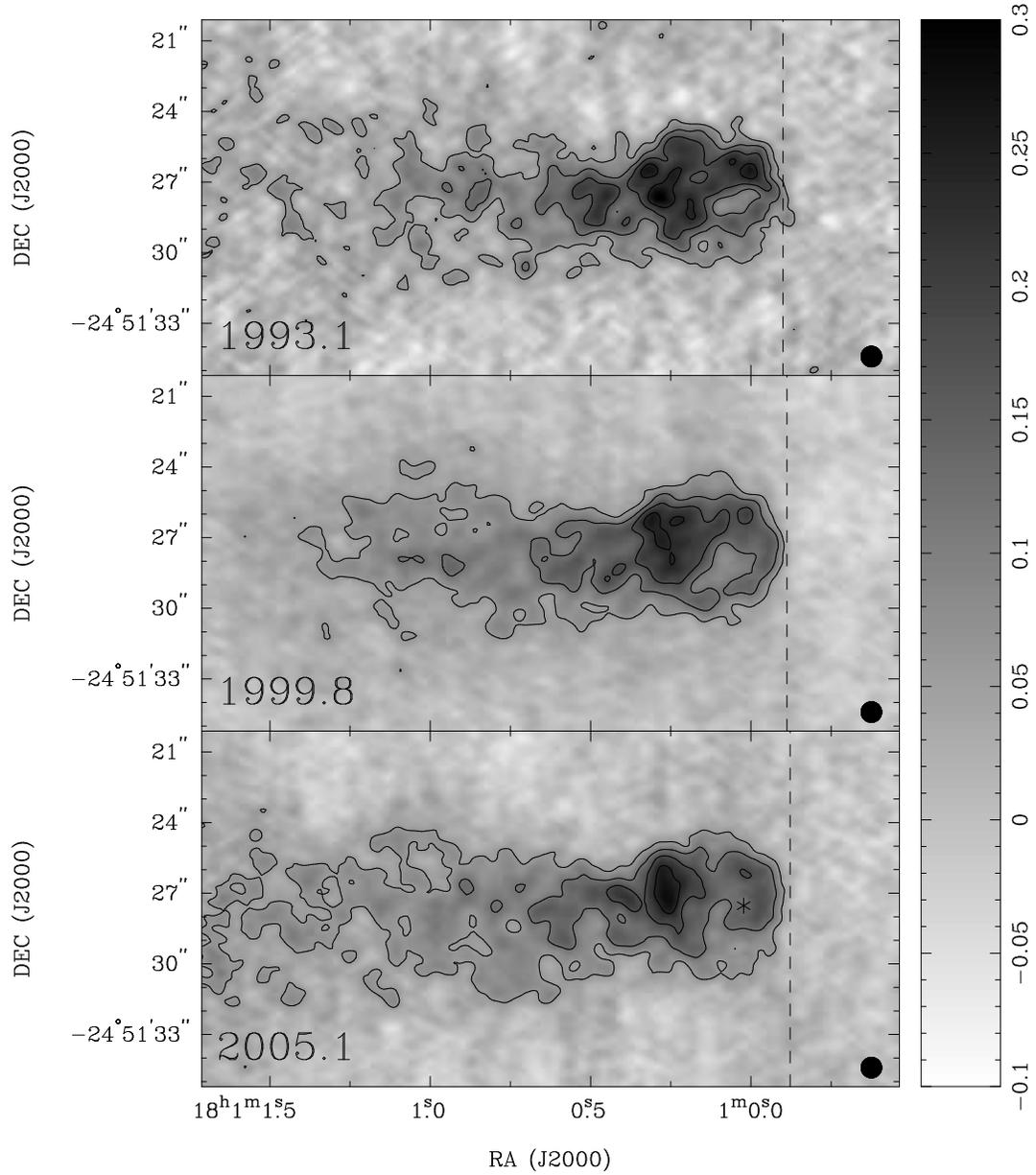}}
\caption{8.5~GHz VLA observations of the western tip of 
PWN~\pwn\ at three epochs
spread over 12~years.  In each panel, the greyscale ranges from --0.1 to
+0.3~mJy~beam$^{-1}$ (as shown in the wedge at right), while the contours
are at levels of 0.6, 1.2, 1.8 and 2.4~mJy~beam$^{-1}$. Each image
has been smoothed to a resolution of FWHM $0\farcs85$, as shown by the
circle at the bottom right of each panel. The position of PSR~\psr\
at epoch 2002.2, as given by TBG02, is marked by an asterisk in the
bottom panel, and has an uncertainty
much smaller than the size of the symbol.
The vertical dashed line shows the expected relative shift
for westward proper motion of 24.8~mas~yr$^{-1}$ (corresponding to the
5-$\sigma$ upper limit previously determined by GF00).
Note that the first epoch has poorer sensitivity than the second and third
observations.}
\label{fig_pwn}
\end{figure}

Analysis was carried out in the MIRIAD package \citep{sk03}.  Data
from the three epochs were reduced in almost identical fashion,
making allowances for the slightly different correlator mode used
in the 1993 data.  The data were edited, calibrated, and imaged
using square pixels of size 50~mas $\times$ 50~mas.  The fields were
then deconvolved using CLEAN and smoothed to a common resolution
of $0\farcs85$. The resulting images, shown in Figure~\ref{fig_pwn},
suggest that there have been slight structural changes between epochs.

The vertical dashed lines in Figure~\ref{fig_pwn} demonstrate that motion
at the 5-$\sigma$ limit of GF00 could now have been easily discerned,
and that any change in the position of the leading edge of the PWN is
well below this level.  We have quantified this result by measuring
the shift between epochs using the MIRIAD task IMDIFF. IMDIFF finds the
shift that minimizes the RMS of intensity fluctuations in the resulting
difference map, employing cubic convolution interpolation to calculate
shifts of a non-integer pixel number \citep{pow64}.

Using the approach described by GF00, the shift
that we determine between the 1993 and 2005 epochs 
is $71\pm24$~mas in a westward direction. 
However, this does not take into account any systematic
errors present in this shift determination, nor does
it use the combination of all three epochs. We have
thus determined the motion of the PWN by
measuring the shift between all three possible pairs
of epochs using IMDIFF.
To characterize the
statistical error in the reported shifts, for each pair a series
of phase shifts were applied to the $u-v$ visibility of one epoch,
resulting in images which had been shifted east or west by 0, 0.4, 0.8,
1.2, $\ldots$, 4.0~pixels. This allowed us to probe the systematic
errors introduced by IMDIFF for non-integer pixel shifts of a faint
extended source in the presence of noise.  To minimize the effects
of any internal structural change in the PWN between epochs, the
brightness distribution above a surface brightness of
70~mJy~beam$^{-1}$ was pegged at 70~mJy~beam$^{-1}$, 
giving the nebula a largely uniform intensity
out to its periphery.  For each pair of epochs and each phase shift,
IMDIFF was then run using 16 different sets of inputs, produced by
alternating between four different spatial
windows,
by switching between the choice of epoch used as the reference
image, and  constraining IMDIFF to find a shift only in
R.A., or in both R.A.\ and Decl. The ensemble of outputs from the
combinations of these options allowed us to characterize, for each
input trial shift, a mean fitted shift in the R.A.\ direction and
its systematic error.  The recovered shifts were then plotted against
the input shifts and fit with a weighted linear least-squares model.
The ordinate intercept then represented the best-fit measured motion
between each pair of epochs. The statistical error on this estimate
was determined by calculating the standard error in the mean amongst
the residuals between input and mean output values for the 21 trial
shifts for each pair. The statistical and mean systematic errors
were then combined in quadrature to give the final error estimate
for each pair.  Other sources of error, such as phase transferral
and calibration error, are negligible compared to these effects and
have not been incorporated in this analysis.

We then examined the correlation between shifts between each of the
three epoch pairs  and the time separation between these pairs. If
steady westward motion was detected, we would have expected these two quantities
to correlate, but the data show no such trend.  We conclude that
despite our best efforts to minimize the effects of structural
changes in the nebula between epochs, the underlying proper motion
is too small in magnitude to be seen in the presence of the
systematic errors introduced by these changes.  We have correspondingly
determined an upper limit on the source's motion by calculating the
$\chi^2$ of the best linear fit to the data, and then finding the
larger of the two slopes for which $\chi^2$ increases by 5-$\sigma$
over this best-fit value.

Through this approach, we find an upper limit on westward proper
motion of \pwn\ of $\mu <  13.9$~mas~yr$^{-1}$. This and previous
estimates are shown in Table~\ref{tab_results}: it can be seen that
this new result is nearly a factor of two more constraining than
existing measurements.  For $\Theta = 16\farcm1 - 20\farcm6$ as
assumed above, the 5-$\sigma$ lower limit on the system's age is
$t_p =\Theta/\mu > 69-88$~kyr (see Table~\ref{tab_results}).

Note that because the X-ray and radio morphologies of  \pwn\ suggest
proper motion purely in a westward direction, we have not included
any shift in Decl.\ in our final proper motion estimates. In any case,
the derived shift between each pair of  epochs is consistent with zero
proper motion in Decl.

Assuming a distance to the system of 5~kpc, we can infer
an upper limit on the projected westward velocity for PSR~\psr\ of
340~\kms.  This limit is consistent with the observed range of
motions for other pulsars associated with SNRs \citep[see Table 6
of][]{hllk05}, and also with the overall projected velocity
distribution of the young pulsar population \citep{acc02,hllk05}.

\section{Discussion}
\label{sec_disc}

Since we have used the motion of PWN~\pwn\ as a proxy for that of PSR~\psr,
the two objects must move together rigidly for our limit to be
applicable to the pulsar.  As discussed by GF00, the pulsar
could be moving faster than our upper limit
if the stand-off distance between the pulsar and the head
of the bow shock were steadily decreasing with time, as might be
produced by the system encountering a sudden increase in density.
However, this is unlikely to be the case since, if anything, the
data in Figure~\ref{fig_pwn} 
suggest that emission from the leading edge of the bow shock
has slightly faded in brightness 
as a function of time. If this is na\"ively interpreted as
corresponding to a reduction in ambient density, then the stand-off
distance could even be increasing with time, further
strengthening our upper limit
on the pulsar's motion.  In any case the
direct observations of the pulsar reported by TBG02 rule out a large
disparity between the proper motions of the pulsar and its PWN.

The angular proximity and consistent distance estimates for
PSR~\psr\ and SNR~\snr\ obviously demand
that the possibility of a physical association be considered. Further
evidence strengthening the case for a genuine association are the
high spin-down luminosity of the pulsar, the fact that the PWN's
cometary tail points
back toward the SNR's interior
(seen in Figs.~\ref{fig_snr} and~\ref{fig_pwn}),
and the edge-brightening and flatter
spectral index of SNR~\snr\ along its western rim near the pulsar
\citep[Fig.~\ref{fig_snr};][]{fkw94}. 
In the absence of any constraints on
the PWN's proper motion or the pulsar's
age, the simplest interpretation
would seem to be that the pulsar was born
reasonably close to the SNR's geometric center, has then moved
outward, overtaken the SNR, and in the process has re-energized the
shell's emission with its relativistic outflow of particles and
magnetic fields \citep{sfs89}.
However, in \S\ref{sec_obs} we have derived $t_p > 69-88$~kyr if
the pulsar was born near the SNR's center.  Since $\tau_c =
15$~kyr~$\ll t_p$, the picture proposed above is problematic.

In the following discussion, we consider three possible solutions\footnote{We
acknowledge the existence of a variety of other explanations
that have been proposed for PSR~\psr\ and SNR~\snr,
such as have been discussed by \cite{kun92b}, \cite{ist94}, \cite{mlr01}
and \cite{sx03}.
However, here we focus on three
simple possibilities which
can potentially be compared using available data.} to this
difficulty: (1) the pulsar and the SNR are physically associated with an
age $t_p \approx \tau_c$, but the pulsar was born substantially
offset from the SNR's
geometric center; (2) the pulsar and SNR are physically
associated, but $t_p \gg \tau_c$; or (3) the two objects
have no physical association. We conclude by considering the
nature of \pwn, which we argue provides additional clues to distinguish
between these possibilities.

\subsection{Association Involving a Young Pulsar}
\label{sec_offc}

In the case of a physical association, but with a supernova explosion
substantially offset from the geometric center of the SNR, our upper limit
on $t_p$ can potentially be made consistent with the characteristic age.
If the blast center of
the SNR were considerably closer to the current location of the pulsar
than the separation of $\Theta = 16\farcm1 - 20\farcm6$ assumed earlier,
the predicted pulsar motion would be lower, and possibly could brought
below the upper limit in Table~\ref{tab_results}.

A reasonable mechanism through which this large offset could have
occurred is if the supernova occurred in the stellar wind bubble of a
moving progenitor star \citep{gva02}.  Expansion of a SNR into such a
bubble could produce a 
remnant whose blast site is substantially
separated from the geometric center \citep{gva04b}. In such a case,
the pulsar could have been born inside the SNR but quite close to its
present location, and the predicted proper motion would then be very low.

This explanation was viable for the upper limits obtained by GF00 and
TBG02. However, for our new, more constraining proper motion measurement,
the pulsar would need to have been born on the rim of or even outside
the SNR to satisfy the requirement that $t_p \approx \tau_c$.  While it
is possible that the progenitor star was in the process of escaping
its own wind bubble when the supernova occurred, such a system would produce
an SNR significantly elongated and distorted along the east-west
axis corresponding to the pulsar's motion
\citep{rtfb93,bd94}, not
consistent with the SNR morphology seen here.

Furthermore, since the pulsar was kicked randomly away from its
birthplace, a wide separation  between the neutron star birth site and
the SNR's geometric center implies that there should most likely be
a significant misalignment between the pulsar's direction of motion
(as inferred from the PWN's morphology) and the vector joining the
SNR's center to the pulsar's current position \citep{gva02,mgb+02}.
Specifically, for
an explosion occurring near the edge of the cavity,
there is a $\approx90\%$ probability that the
misalignment between the 
projections of these two vectors will be 
larger than the $\approx20^\circ$ observed.
The mild misalignment between the
projections of these two vectors therefore makes 
a scenario in which the explosion
was substantially off-center relatively unlikely.

\subsection{Association Involving an Old Pulsar}
\label{sec_old}

As suggested by GF00 and \cite{gva04b}, the case for a pulsar / SNR
association can be made if the characteristic age is a significant
underestimate of the system's true age.  The solution proposed by GF00
predicts an evolved system consisting of an old SNR and a slow-moving
pulsar, for which $t_p = 93-170$~kyr and $\mu=7-10$~mas~yr$^{-1}$,
consistent with the upper limits found here.  However, we then need to
explain why $t_p \gg \tau_c$.

To consider this possibility, we must reconsider the assumptions
underlying Equation~(\ref{eqn_age}).  The braking
index, $n$,  is defined by the equation
$\dot{\nu}=-K\nu^{n}$, where $\nu \equiv 1/P$. 
For spin-down via a magnetic dipole {\em in vacuo},
we expect $n=3$.
Following \cite{br88}, we separately define
the ``deceleration parameter'',
$\tilde{n} \equiv \nu \ddot{\nu}/ \dot{\nu}^2$.
If $K$ and $n$ are both constant,
then $\tilde{n} = n$.

There are two situations in which the characteristic age, $\tau_c
= P/2\dot{P}$, can underestimate the true age as required here.
The first possibility is that $P \gg P_0$ and $K$ is constant, but
$n = \tilde{n}  < 3$.  For PSR~\psr, Equation~(\ref{eqn_age}) can
only yield $t_p > 69$~kyr for $P_0 < 13$~ms, and even then only for
$\tilde{n} \approx 1$.  In contrast, measurements of initial spin
in most other systems suggest $P_0 \ga 20$~ms \citep{mgb+02,klh+03},
while in the few cases where $\tilde{n}$ has been determined, it
is found that $1.4 < \tilde{n} < 2.9$ (Livingstone \etal\ 2005,
2006, and references therein\nocite{lkg05,lkgk06}).  It thus seems
unlikely that a viable set of parameters can describe the system
via Equation~(\ref{eqn_age}).

The alternative is that $K$ is not constant.  In this case,
Equation~(\ref{eqn_age}) no longer holds, and $\tilde{n} \neq n$
\cite[see][]{br88}.  Either a changing magnetic field or a changing
moment of inertia can cause $K$ to vary with time \citep[see
e.g.,][]{cam96b}.  In this case, the pulsar spin-down is uncoupled
from the star's age, and rather traces the time scale on which $K$
evolves.

While it seems unlikely that the star's moment of inertia could
evolve substantially on this time-scale, the growth of the surface
magnetic field  is a possibility, as may result from thermoelectric
instabilities in the crust or diffusion of magnetic flux from the
star's interior \citep{bah83,rzc98,kg01,lz04}.  
If we assume $n=3$ but
$\tilde{n} < 3$, then 
$\tau_c$ underestimates the true age,
and the surface magnetic field, $B$, grows with time \citep{lyn04}.
Indeed $\tilde{n} < 3$ is observed for all six pulsars
in which this quantity has been accurately determined,
while $dB/dt > 0$ is suggested in long-term timing signatures
of several other pulsars \citep{lpgc96,smi99b,lyn04}.
While $\tilde{n}$ has not been
directly measured for PSR~\psr, we can write
as an order of magnitude estimate that the time scale
for field growth has been
$B/\dot{B} \sim \tau_c \approx 15$~kyr.

We further note that this does not require us to invoke a unique
effect to explain this specific pulsar / SNR association.  Even
though $\tau_c > t_p$ for many other pulsars in SNRs, these results
are still consistent with $\tilde{n} < 3$ \citep{krv+01,mgb+02}.
Gradual magnetic field growth may thus be widespread, but is only  noticeable
in a system such as the Duck for which its comparatively large age
has allowed this effect to accumulate.

It is important to acknowledge that the large implied age ($t_p > 69$~kyr)
is beyond what is expected for the observable lifetime of a SNR,
which is typically $\sim20-60$~kyr \citep{bgl89,fgw94}.  Thus even
if magnetic field growth can explain the discrepancy between the
pulsar's characteristic age and that inferred from proper motion, the
large implied age is also a potential issue for SNR~\snr.  
However, the faint eastern rim of the SNR is consistent with what
is seen for older, undisturbed shells, and may not have been easily
identified on its own if not for the brighter western side.  Thus
the large age required for the SNR is not inconsistent with the
appearance of this half of the SNR.

\begin{figure}
\centerline{\includegraphics[width=\textwidth]{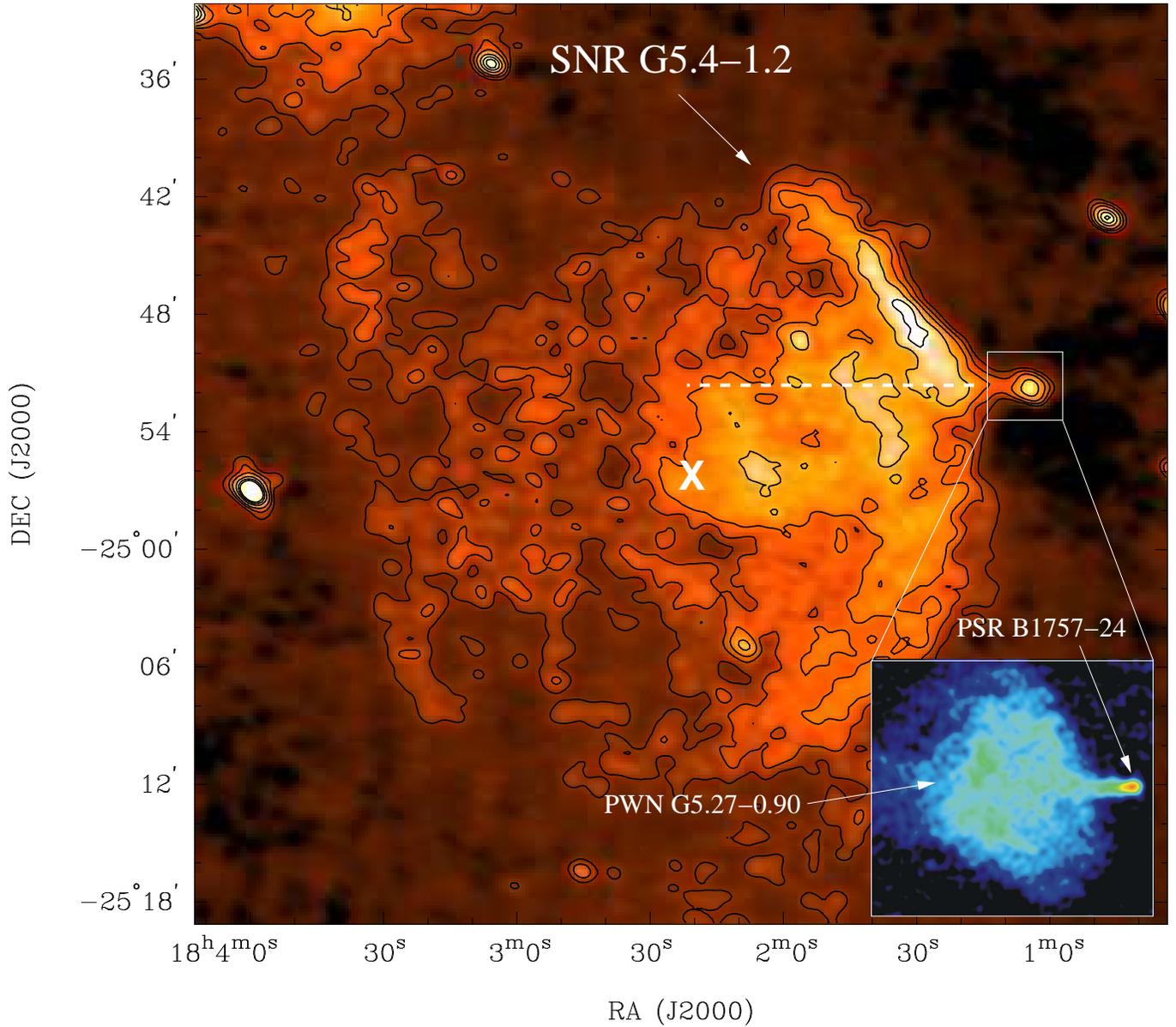}}
\caption{VLA observations of ``the Duck'', with the SNR, PWN and pulsar
indicated. The main panel
shows a 327-MHz image of SNR~\snr\ at a resolution
of $55''\times41''$, adapted from GF00.  Contours are at levels
of 10, 25, 50 and 100~mJy~beam$^{-1}$, and the peak intensity is
150~mJy~beam$^{-1}$. The
cross shows the approximate location of the center of
SNR~\snr, while the horizontal line indicates the inferred direction of
motion for the pulsar as implied by the morphology of \pwn.
The inset shows a 8.5-GHz image of PWN~\pwn, covering a
$3\farcm5\times3\farcm5$ field at a resolution of $\sim1''$ (image
courtesy of NRAO/AUI and Dale A.\ Frail).}
\label{fig_snr}
\end{figure}

Furthermore, if the pulsar and SNR are physically associated, the
discrepancy between the system's inferred age and the expected lifetime
for observable SNRs can be resolved by the argument that the pulsar
is re-energizing the SNR as it passes through it \citep{sfs89}.
The re-energization hypothesis predicts that brighter emission with
a flatter spectrum should be seen along the western edge of SNR,
and that these effects should peak where the pulsar would have
crossed the rim. Indeed these phenomena are both observed
\citep[Fig.~\ref{fig_snr};][]{bh85,fkw94}. However, a requirement
of this theory is that synchrotron-emitting particles must diffuse
sufficiently rapidly away from the pulsar around the rim of the
shell to produce the observed region of apparent interaction
\citep{vag02}.  We can quantify this as follows.  If $T$ is the
time elapsed since the pulsar first began to interact with
the SNR shell and $X$ is
the distance which is traveled by particles  along the shell's circumference
after injection by the pulsar wind, then we require $X =
(2\kappa_{\rm Duck} T)^{1/2}$, where $\kappa_{\rm Duck}$ is the
diffusion coefficient of synchrotron emitting particles as they
move from the pulsar around the rim. From Figure~\ref{fig_snr}
we estimate $X \approx 10$~pc and $T \sim 0.1 t_p \ga
7$~kyr.  We thus require $\kappa_{\rm Duck} \sim
2\times10^{27}$~cm$^2$~s$^{-1}$ to explain the SNR's appearance.

We can determine if this inferred diffusion coefficient is reasonable
by considering the magnetic field configuration through which
particles must propagate.  If the SNR is in the radiative phase as
argued by GF00, then we expect strong compression at the SNR shock,
and hence a magnetic field which runs parallel to the shell. Indeed
radio polarization measurements clearly demonstrate this geometry
to be present along the SNR's western rim \citep{mch92}. The mean
free path in directions parallel to the magnetic field lines is larger than the
electron gyroradius by a factor $\eta_{\rm Duck} = (\delta B_0 / B_0)^{-2}$,
where $B_0$ is the ambient magnetic field strength 
in the SNR shell and $\delta B_0$ is the
amplitude of turbulent fluctuations in $B_0$ \citep{jok87,abr94}.
We can thus write
$\kappa_{\rm Duck} = \eta_{\rm Duck} \kappa_B$, where $\kappa_B$ is
the standard Bohm diffusion coefficient, with value $\kappa_B \approx
1\times 10^{23} (B_0/{\rm \mu G})^{-3/2}$~cm$^2$~s$^{-1}$ for
the 327-MHz image shown in Figure~\ref{fig_snr} \citep{vag02}. We 
then require $\eta_{\rm Duck} \sim 2 (B_0/{\rm \mu G})^{3/2}\times10^4$.
In comparison, \cite{abr94} find
$\eta_{\rm young} \la 2000$ for young SNRs which have enhanced turbulence,
but $\eta_{\rm ISM} \ga 10^5$ for the ISM. If we assume a
field strength $B_0 \sim 5-10$~$\mu$G,  we thus
find $\eta_{\rm Duck} \sim \eta_{\rm ISM}$, consistent
with SNR~\snr\ being an old remnant with reduced
turbulent amplitude which is merging into the ISM
\citep[see also][]{mr94}.
We conclude that the re-energization hypothesis
is consistent with the expected rapid diffusion of particles
from the pulsar around the shell rim.

As a final note, we point out that even if the pulsar is old, a small
offset between the supernova explosion site and the SNR's geometric
center is still required.  Specifically, the trajectory of the
pulsar as inferred from the cometary tail of the PWN passes north
of the SNR's geometric center, indicating that the blast and geometric
centers do not coincide \citep[Fig.\protect\ref{fig_snr};][]{fkw94}.
\cite{fkw94} proposed that a gradient in the density of the ambient
interstellar medium (ISM) into which the SNR is expanding could
produce an asymmetric expansion, indeed resulting in a small offset
between the explosion site and the SNR's center. Alternatively,
\cite{gva04b} has
proposed that the misalignment
between the pulsar's expected and inferred trajectories
could be due to a progenitor star which explodes slightly
offset from the center of its wind-blown bubble
(as already discussed in \S\ref{sec_offc}, but for the case of a young pulsar
with a subtantial offset from the bubble's center).
This again leads to the conclusion that the pulsar's true age is
considerably in excess of its characteristic age \citep[see][]{gva04b}.

\subsection{Chance Alignment}

The alternative simple explanation proposed by TBG02\nocite{tbg02}
is that SNR~\snr\ and PSR~\psr\ have no physical connection.
Given the relatively high density on
the sky of both pulsars and SNRs in the inner Galactic plane, it is
reasonable that two such objects could appear near each
other in projection \citep{gj95c}.  
\HI\ absorption is only able to provide lower
limits on the distances to the SNR and PWN \citep{fkw94}, 
while the pulsar's distance
estimate as derived from its dispersion measure is model dependent and
comes with significant associated uncertainties \cite[see discussion
by][]{cl02}.  Thus, although the
distances to the two objects are consistent, TBG02 point out that this
is not necessarily strong evidence in favor of an association.

Previous discussions had focused on the flatter spectrum and brighter
emission  of the SNR as evidence for an association, as described
above.  However, TBG02 point out that variations in spectral index
and asymmetries in brightness are both common in SNRs. In particular,
many SNRs are brighter on the side closest to the Galactic Plane,
as observed here.  While these are valid arguments, we note that
in most other SNRs in which spectral index variations are observed,
these changes are spread randomly over the SNR, rather than
concentrated in a particular region \citep{ar93}. More
specifically, the systematic trend toward flatter spectra as one
gets closer to the pulsar position around the rim of \snr\ is
difficult to explain if the SNR and pulsar are unassociated.

\subsection{The Nature of \pwn}

An important additional aspect of this discussion is the nature of \pwn.
As TBG02\nocite{tbg02} note, the very tight angular coincidence of
PSR~\psr\ and \pwn, along with the cometary morphology of the latter,
make it virtually certain that at least these two objects are associated.

It is important to bear in mind  that Figure~\ref{fig_pwn} shows only
the westernmost extent of this source. The inset to Figure~\ref{fig_snr}
demonstrates that eastward of the pulsar, this structure broadens into a
larger nebula, approximately $100''\times100''$ in extent.  TBG02 interpret this
overall morphology as a ``Crab-like'' PWN. In their model, the pulsar
was born at the center of this larger nebula, and is now moving away
from this site.
However, there are difficulties with this interpretation.
For the pulsar to have escaped from its own wind-driven bubble, the
expansion speed of the PWN must have fallen well below the pulsar's
space velocity. This process can only occur in two scenarios.

The first possibility
is that as a PWN expands into the freely expanding ejecta of its
associated SNR, it will eventually collide with the SNR reverse
shock \citep{rc84,vagt01,bcf01}. The resulting interaction can
reduce and even reverse the expansion of the PWN.  Combined with
the pulsar's ballistic motion, this compression of the PWN produces
a morphology in which the pulsar is at the tip of a trail of
radio/X-ray emission, connected to a larger ``relic PWN''. This
scenario was invoked by \cite{vdk04} to explain the morphologies
of the PWNe seen in SNRs N157B and G327.7--1.1.  The difficulty
with this interpretation here is that to have produced a reverse shock,
the SNR needs to have interacted with and swept up a significant
amount of interstellar gas.  The fact that we do not see any radio
emission from an associated SNR makes this interpretation
problematic.

The alternative possibility 
is that no significant outer blast wave
was produced, as may be the case for the Crab Nebula
and for 3C~58 \citep{shvm04,sgs06}. 
The PWN then interacts directly with the ISM, and
has decelerated sufficiently as it sweeps up ambient gas that the pulsar
is now able to overtake it. 
In this case we can write \citep{cmw75}:
\begin{equation}
R_{\rm PWN} \approx \left(\dot{E}/\rho\right)^{1/5} t^{3/5} ,
\end{equation}
where $R_{\rm PWN}$ is the radius of the ``relic'' component of the
nebula at time $t$, $\rho$ is the ambient mass density, and where we
have assumed that the pulsar blows a steady wind of luminosity $\dot{E}$
into a uniform surrounding medium. For
$t = \tau_c \approx 15$~kyr, $\dot{E} = 2.6\times10^{36}$~ergs~s$^{-1}$
and $R_{\rm PWN} \approx 1$~pc, we then find an ambient number
density $n_0 \sim 500$~cm$^{-3}$. 

This high value can be ruled out by the morphology of the bow-shock
component of the PWN. As discussed by GF00, the stand-off
distance between the pulsar position and the apex of the bow shock is
$r_w \approx 0.04$~pc. This position is set by pressure balance between
the pulsar wind and the ram pressure of the surroundings, $\dot{E}/4\pi
r_w^2 c \approx \rho V^2$, where $V$ is the pulsar space velocity. The
left hand term yields a pressure $\sim5\times10^{-10}$~ergs~cm$^{-3}$. For
the density inferred above, we then find $V \sim 10$~\kms. Not only is
this velocity more than an order of magnitude slower than seen for young
pulsars, but the solution is not self-consistent, since in 15~kyr, such
a pulsar would only have traveled $\sim0.1-0.2$~pc and so could not have
moved outside its PWN bubble.

We thus conclude that there is no obvious explanation for
\pwn\ if PSR~\psr\ is $\sim15$~kyr old and is unassociated
with SNR~\snr.

On the other hand, if there is a genuine association between the pulsar
and the SNR,
then \pwn\ could represent the remnants of the interaction
between the pulsar wind and the SNR shell.  \cite{vag+03} consider the
interaction between a high-velocity pulsar and a SNR in the Sedov-Taylor
phase of evolution. They show that as the pulsar crosses the SNR, the drop
in ram pressure from that in the SNR shell to that of the ambient ISM
results in an expansion of the bow-shock structure.  However, if \snr\
is in the pressure-driven ``snowplow'' phase of evolution (as argued by GF00
and as would be typical for the age inferred in Table~\ref{tab_results}),
then the compression ratio of gas swept up by the forward shock will be
much larger than the standard factor of four seen in the Sedov solution
\citep[e.g.,][]{bwbr98}.  Therefore, during the period in
which the pulsar breaks through the rim of the shell, the ram pressure
experienced by the bow shock drops dramatically, which in turn should
produce a sudden, explosive expansion of the pulsar wind nebula.
As the pulsar moves outward, the pulsar wind blows a ``Crab-like''
structure into the ISM, which might possibly correspond to \pwn.
Eventually the pulsar readjusts to its new, lower density environment,
again forming a standard Mach cone, but connected to
the larger relic structure left behind.
This idea needs testing by full hydrodynamic modeling, but
it provides a feasible explanation for the overall morphology of \pwn\
in the scenario in which the pulsar is old and is associated with the SNR.

We note that \cite{gva04b} has proposed an alternate explanation for \pwn,
in which this source is a slowly expanding lobe produced by the collimated
flow of hot gas which follows the pulsar as it punctures the SNR shell. This
similarly requires a physical association between the pulsar and the SNR.

\section{Conclusions}

The upper limit on proper motion we find for the PWN~\pwn\ is the
most restrictive one obtained to date for this system.  We are able
to reject the original expectation that PSR~\psr\ was born
$\sim15$~kyr ago and is associated with SNR~\snr, 
regardless of whether the corresponding supernova occurred at the geometric
center of the SNR, or at a site substantially offset from this.

Two possibilities remain. First, the pulsar and SNR are associated and
share an age $\ga70$~kyr. In this case, the pulsar has a typical
projected space velocity of $\la330$~\kms, which allows it to overtake
its SNR at this stage and drive a bow shock through ambient gas.
The sudden drop in pressure as the pulsar crosses the SNR's radiative
shell might also explain the bulbous component of the PWN~\pwn\ seen
between the pulsar and the SNR.  The pulsar's passage has re-energized
the radio emission from the SNR through rapid diffusion of pulsar wind
particles along magnetic field lines; without this new injection of
particles, the SNR would be much dimmer, would have a steeper spectrum and
generally would be more difficult to detect.  The slight offset between
the pulsar's inferred trajectory and the SNR's geometric center possibly
results from a density gradient into which the SNR is expanding,
or from a slightly offset explosion within a pre-existing cavity.
The only difficulty with this
picture is that the system's age must then be many times larger than
the pulsar's characteristic age, which is not seen in other pulsar /
SNR associations. The proposed explanation is that the surface magnetic
field of the pulsar is at the present epoch growing on a time scale of
$\sim15$~kyr. This effect is consistent with the properties of other
pulsar / SNR associations, and is possibly also being seen in the
long-term timing signatures of several pulsars \citep{lpgc96,lyn04}.

The alternative is that the pulsar is $\sim15$~kyr old as indicated
by its characteristic age, 
and is unassociated with SNR~\snr.  This explanation
eliminates the need to invoke off-center cavity explosions,
re-energized shells or growing magnetic fields to explain the
observations. However, this interpretation offers no easy explanation
for the morphology of \pwn\ which, contrary to the proposal of
TBG02, cannot be easily explained as a relic nebula left behind at the
pulsar's birth site. This model also requires that
the brightened emission and flat spectrum of the SNR near
the pulsar be a coincidence.

Frustratingly, a full understanding of the Duck remains elusive.
However, future VLA or VLBA observations of
PSR~\psr\ should eventually detect motion. Furthermore, an X-ray
detection of \snr\ and of the eastern parts of \pwn\ 
should provide additional constraints on
the properties  of the system 
\citep[see discussion by][]{kggl01},
while continued timing of the pulsar may
be able to indicate the nature of the braking torque and magnetic
field evolution in this source \citep[e.g.,][]{lpgc96}.

Finally, other similar systems might provide additional
clues.  PSR~B1951+32 is clearly in the process of puncturing the
shell of the SNR~CTB~80 \citep{hk88,fss88}, while PSR~J1016--5857
and SNR G284.3--1.8 have been proposed as another such interacting
system \citep{cbm+01}. Over the  ensemble of Galactic pulsars and
SNRs, \cite{sfs89} predict approximately half a dozen systems in
which the pulsar has recently penetrated the SNR shell.  Searches
for further interacting pairs may thus prove fruitful.

\begin{acknowledgements}

We thank Josh Grindlay for useful discussions, and the referee,
Vasilii Gvaramadze, for helpful suggestions.  The National Radio
Astronomy Observatory is a facility of the National Science Foundation,
operated under cooperative agreement by Associated Universities,
Inc. We acknowledge use of the NRAO Image Gallery for the inset to
Figure~\ref{fig_snr}.  J.A.B. and B.M.G. are supported by NASA through
LTSA grant NAG5-13032. S.C. is a Jansky Fellow of the National Radio
Astronomy Observatory.  F.C.  is supported by  the NSF through grant
AST-0507376.

\end{acknowledgements}


\begin{thebibliography}{59}
\expandafter\ifx\csname natexlab\endcsname\relax\def\natexlab#1{#1}\fi

\bibitem[{Achterberg {et~al.}(1994)Achterberg, Blandford, \& Reynolds}]{abr94}
Achterberg, A., Blandford, R.~D., \& Reynolds, S.~P. 1994, A\&A, 281, 220

\bibitem[{Anderson \& Rudnick(1993)}]{ar93}
Anderson, M.~C. \& Rudnick, L. 1993, ApJ, 408, 514

\bibitem[{Arzoumanian {et~al.}(2002)Arzoumanian, Chernoff, \& Cordes}]{acc02}
Arzoumanian, Z., Chernoff, D.~F., \& Cordes, J.~M. 2002, ApJ, 568, 289

\bibitem[{Becker \& Helfand(1985)}]{bh85}
Becker, R.~H. \& Helfand, D.~J. 1985, Nature, 313, 115

\bibitem[{Blandford {et~al.}(1983)Blandford, Applegate, \& Hernquist}]{bah83}
Blandford, R.~D., Applegate, J.~H., \& Hernquist, L. 1983, MNRAS, 204, 1025

\bibitem[{Blandford \& Romani(1988)}]{br88}
Blandford, R.~D. \& Romani, R.~W. 1988, MNRAS, 234, 57P

\bibitem[{Blondin {et~al.}(2001)Blondin, Chevalier, \& Frierson}]{bcf01}
Blondin, J.~M., Chevalier, R.~A., \& Frierson, D.~M. 2001, ApJ, 563, 806

\bibitem[{Blondin {et~al.}(1998)Blondin, Wright, Borkowski, \&
  Reynolds}]{bwbr98}
Blondin, J.~M., Wright, E.~B., Borkowski, K.~J., \& Reynolds, S.~P. 1998, ApJ,
  500, 342

\bibitem[{Braun {et~al.}(1989)Braun, Goss, \& Lyne}]{bgl89}
Braun, R., Goss, W.~M., \& Lyne, A.~G. 1989, ApJ, 340, 355

\bibitem[{Brighenti \& D'Ercole(1994)}]{bd94}
Brighenti, F. \& D'Ercole, A. 1994, MNRAS, 270, 65

\bibitem[{Camilo(1996)}]{cam96b}
Camilo, F. 1996, in Pulsars: Problems and Progress, {IAU} Colloquium 160, ed.
  S.~Johnston, M.~A. Walker, \& M.~Bailes (San Francisco: Astronomical Society
  of the Pacific), 39

\bibitem[{Camilo {et~al.}(2001)Camilo, Bell, Manchester, Lyne, Possenti,
  Kramer, Kaspi, Stairs, D'Amico, Hobbs, Gotthelf, \& Gaensler}]{cbm+01}
Camilo, F., Bell, J.~F., Manchester, R.~N., Lyne, A.~G., Possenti, A., Kramer,
  M., Kaspi, V.~M., Stairs, I.~H., D'Amico, N., Hobbs, G., Gotthelf, E.~V., \&
  Gaensler, B.~M. 2001, ApJ, 557, L51

\bibitem[{Castor {et~al.}(1975)Castor, McCray, \& Weaver}]{cmw75}
Castor, J., McCray, R., \& Weaver, R. 1975, ApJ, 200, L107

\bibitem[{Caswell {et~al.}(1987)Caswell, Kesteven, Komesaroff, Haynes, Milne,
  Stewart, \& Wilson}]{ckk+87}
Caswell, J.~L., Kesteven, M.~J., Komesaroff, M.~M., Haynes, R.~F., Milne,
  D.~K., Stewart, R.~T., \& Wilson, S.~G. 1987, MNRAS, 225, 329

\bibitem[{{Cordes} \& {Lazio}(2002)}]{cl02}
{Cordes}, J.~M. \& {Lazio}, T.~J.~W. 2002, {astro-ph/0207156}

\bibitem[{Fesen {et~al.}(1988)Fesen, Shull, \& Saken}]{fss88}
Fesen, R.~A., Shull, J.~M., \& Saken, J.~M. 1988, Nature, 334, 229

\bibitem[{Frail {et~al.}(1994{\natexlab{a}})Frail, Goss, \& Whiteoak}]{fgw94}
Frail, D.~A., Goss, W.~M., \& Whiteoak, J. B.~Z. 1994{\natexlab{a}}, ApJ, 437,
  781

\bibitem[{Frail {et~al.}(1994{\natexlab{b}})Frail, Kassim, \& Weiler}]{fkw94}
Frail, D.~A., Kassim, N.~E., \& Weiler, K.~W. 1994{\natexlab{b}}, AJ, 107, 1120

\bibitem[{Frail \& Kulkarni(1991)}]{fk91}
Frail, D.~A. \& Kulkarni, S.~R. 1991, Nature, 352, 785

\bibitem[{{Gaensler} \& {Frail}(2000)}]{gf00}
{Gaensler}, B.~M. \& {Frail}, D.~A. 2000, Nature, 406, 158 (GF00)

\bibitem[{Gaensler \& Johnston(1995)}]{gj95c}
Gaensler, B.~M. \& Johnston, S. 1995, MNRAS, 277, 1243

\bibitem[{Gvaramadze(2002)}]{gva02}
Gvaramadze, V.~V. 2002, in Neutron Stars in Supernova Remnants, ed. P.~O. Slane
  \& B.~M. Gaensler (San Francisco: Astronomical Society of the Pacific),
  23--26

\bibitem[{Gvaramadze(2004)}]{gva04b}
Gvaramadze, V.~V. 2004, A\&A, 415, 1073

\bibitem[{Hester \& Kulkarni(1988)}]{hk88}
Hester, J.~J. \& Kulkarni, S.~R. 1988, ApJ, 331, L121

\bibitem[{Hobbs {et~al.}(2005)Hobbs, Lorimer, Lyne, \& Kramer}]{hllk05}
Hobbs, G., Lorimer, D.~R., Lyne, A.~G., \& Kramer, M. 2005, MNRAS, 360, 974

\bibitem[{Istomin(1994)}]{ist94}
Istomin, Y.~N. 1994, A\&A, 283, 85

\bibitem[{Jokipii(1987)}]{jok87}
Jokipii, J.~R. 1987, ApJ, 313, 842

\bibitem[{Kaspi {et~al.}(2001{\natexlab{a}})Kaspi, Gotthelf, Gaensler, \&
  Lyutikov}]{kggl01}
Kaspi, V.~M., Gotthelf, E.~V., Gaensler, B.~M., \& Lyutikov, M.
  2001{\natexlab{a}}, ApJ, 562, L163

\bibitem[{Kaspi {et~al.}(2001{\natexlab{b}})Kaspi, Roberts, Vasisht, Gotthelf,
  Pivovaroff, \& Kawai}]{krv+01}
Kaspi, V.~M., Roberts, M. S.~E., Vasisht, G., Gotthelf, E.~V., Pivovaroff, M.,
  \& Kawai, N. 2001{\natexlab{b}}, ApJ, 560, 371

\bibitem[{Konenkov \& Geppert(2001)}]{kg01}
Konenkov, D. \& Geppert, U. 2001, MNRAS, 325, 426

\bibitem[{{Kramer} {et~al.}(2003){Kramer}, {Lyne}, {Hobbs}, {L{\" o}hmer},
  {Carr}, {Jordan}, \& {Wolszczan}}]{klh+03}
{Kramer}, M., {Lyne}, A.~G., {Hobbs}, G., {L{\" o}hmer}, O., {Carr}, P.,
  {Jordan}, C., \& {Wolszczan}, A. 2003, ApJ, 593, L31

\bibitem[{Kundt(1992)}]{kun92b}
Kundt, W. 1992, Ap\&SS, 190, 159

\bibitem[{Lin \& Zhang(2004)}]{lz04}
Lin, J.~R. \& Zhang, S.~N. 2004, ApJ, 615, L133

\bibitem[{{Livingstone} {et~al.}(2005){Livingstone}, {Kaspi}, \&
  {Gavriil}}]{lkg05}
{Livingstone}, M.~A., {Kaspi}, V.~M., \& {Gavriil}, F.~P. 2005, ApJ, 633, 1095

\bibitem[{Livingstone {et~al.}(2006)Livingstone, Kaspi, Gotthelf, \&
  Kuiper}]{lkgk06}
Livingstone, M.~A., Kaspi, V.~M., Gotthelf, E.~V., \& Kuiper, L. 2006, ApJ,
  submitted (astro-ph/0601530)

\bibitem[{Lorimer \& Kramer(2005)}]{lk05}
Lorimer, D.~R. \& Kramer, M. 2005, Handbook of Pulsar Astronomy (Cambridge
  University Press)

\bibitem[{Lyne(2004)}]{lyn04}
Lyne, A.~G. 2004, in Young Neutron Stars and Their Environments, {IAU}
  Symposium 218, ed. F.~Camilo \& B.~M. Gaensler (San Francisco: Astronomical
  Society of the Pacific), 257--260

\bibitem[{Lyne {et~al.}(1996)Lyne, Pritchard, Graham-Smith, \& Camilo}]{lpgc96}
Lyne, A.~G., Pritchard, R.~S., Graham-Smith, F., \& Camilo, F. 1996, Nature,
  381, 497

\bibitem[{{Manchester} {et~al.}(2005){Manchester}, {Hobbs}, {Teoh}, \&
  {Hobbs}}]{mhth05}
{Manchester}, R.~N., {Hobbs}, G.~B., {Teoh}, A., \& {Hobbs}, M. 2005, AJ, 129,
  1993

\bibitem[{Manchester {et~al.}(1991)Manchester, Kaspi, Johnston, Lyne, \&
  D'Amico}]{mkj+91}
Manchester, R.~N., Kaspi, V.~M., Johnston, S., Lyne, A.~G., \& D'Amico, N.
  1991, MNRAS, 253, 7P

\bibitem[{{Marsden} {et~al.}(2001){Marsden}, {Lingenfelter}, \&
  {Rothschild}}]{mlr01}
{Marsden}, D., {Lingenfelter}, R.~E., \& {Rothschild}, R.~E. 2001, ApJ, 547,
  L45

\bibitem[{Migliazzo {et~al.}(2002)Migliazzo, Gaensler, Backer, Stappers,
  van~der Swaluw, \& Strom}]{mgb+02}
Migliazzo, J.~M., Gaensler, B.~M., Backer, D.~C., Stappers, B.~W., van~der
  Swaluw, E., \& Strom, R.~G. 2002, ApJ, 567, L141

\bibitem[{Milne {et~al.}(1992)Milne, Caswell, \& Haynes}]{mch92}
Milne, D.~K., Caswell, J.~L., \& Haynes, R.~F. 1992, MNRAS, 255, 707

\bibitem[{Moffett \& Reynolds(1994)}]{mr94}
Moffett, D.~A. \& Reynolds, S.~P. 1994, ApJ, 425, 668

\bibitem[{Powell(1964)}]{pow64}
Powell, M. J.~D. 1964, The Computer Journal, 7, 303

\bibitem[{Reynolds \& Chevalier(1984)}]{rc84}
Reynolds, S.~P. \& Chevalier, R.~A. 1984, ApJ, 278, 630

\bibitem[{R\'{o}\.{z}yczka {et~al.}(1993)R\'{o}\.{z}yczka, Tenorio-Tagle,
  Franco, \& Bodenheimer}]{rtfb93}
R\'{o}\.{z}yczka, M., Tenorio-Tagle, G., Franco, J., \& Bodenheimer, P. 1993,
  MNRAS, 261, 674

\bibitem[{Ruderman {et~al.}(1998)Ruderman, Zhu, \& Chen}]{rzc98}
Ruderman, M., Zhu, T., \& Chen, K. 1998, ApJ, 492, 267

\bibitem[{Sault \& Killeen(2004)}]{sk03}
Sault, R.~J. \& Killeen, N. E.~B. 2004, The Miriad User's Guide (Sydney:
  Australia Telescope National Facility),
  (http://www.atnf.csiro.au/computing/software/miriad/)

\bibitem[{Seward {et~al.}(2006)Seward, Gorenstein, \& Smith}]{sgs06}
Seward, F.~D., Gorenstein, P., \& Smith, R.~K. 2006, ApJ, 636, 873

\bibitem[{Shi \& Xu(2003)}]{sx03}
Shi, Y. \& Xu, R.~X. 2003, ApJ, 596, L75

\bibitem[{Shull {et~al.}(1989)Shull, Fesen, \& Saken}]{sfs89}
Shull, J.~M., Fesen, R.~A., \& Saken, J.~M. 1989, ApJ, 346, 860

\bibitem[{Slane {et~al.}(2004)Slane, Helfand, van~der Swaluw, \&
  Murray}]{shvm04}
Slane, P., Helfand, D.~J., van~der Swaluw, E., \& Murray, S.~S. 2004, ApJ, 616,
  403

\bibitem[{Smith(1999)}]{smi99b}
Smith, F.~G. 1999, in Pulsar Timing, General Relativity and the Internal
  Structure of Neutron Stars, ed. Z.~Arzoumanian, F.~Van~der Hooft, \& E.~P.~J.
  van~den Heuvel (Amsterdam: Koninklijke Nederlandse Akademie van
  Wetenschappen), 151--155

\bibitem[{{Thorsett} {et~al.}(2002){Thorsett}, {Brisken}, \& {Goss}}]{tbg02}
{Thorsett}, S.~E., {Brisken}, W.~F., \& {Goss}, W.~M. 2002, ApJ, 573, L111 (TBG02)

\bibitem[{van~der Swaluw {et~al.}(2002)van~der Swaluw, Achterberg, \&
  Gallant}]{vag02}
van~der Swaluw, E., Achterberg, A., \& Gallant, Y.~A. 2002, in Neutron Stars in
  Supernova Remnants, ed. P.~O. Slane \& B.~M. Gaensler (San Francisco:
  Astronomical Society of the Pacific), 135--140

\bibitem[{van~der Swaluw {et~al.}(2003)van~der Swaluw, Achterberg, Gallant,
  Downes, \& Keppens}]{vag+03}
van~der Swaluw, E., Achterberg, A., Gallant, Y.~A., Downes, T.~P., \& Keppens,
  R. 2003, A\&A, 397, 913

\bibitem[{van~der Swaluw {et~al.}(2001)van~der Swaluw, Achterberg, Gallant, \&
  T\'{o}th}]{vagt01}
van~der Swaluw, E., Achterberg, A., Gallant, Y.~A., \& T\'{o}th, G. 2001, A\&A,
  380, 309

\bibitem[{van~der Swaluw {et~al.}(2004)van~der Swaluw, Downes, \&
  Keegan}]{vdk04}
van~der Swaluw, E., Downes, T.~P., \& Keegan, R. 2004, A\&A, 420, 937

\end{thebibliography}

\begin{deluxetable}{cccc}[b!]
\tablecaption{Proper Motion, Velocity and Age Limits on \pwn\ and \psr.\label{tab_results}}
\tablehead{ & GF00 & TBG02 & This paper \\}
\startdata
Proper motion (westward; mas yr$^{-1}$) & $<24.8$ & $<37$  & $<13.9$ \\
Projected velocity (westward; km~s$^{-1}$)\tablenotemark{a} & $<590$  &
$<880$ & $<340$ \\
Age~(kyr)\tablenotemark{b} & $>39-49$ & $>26-33$  & $>69-88$ \\
\enddata

\tablecomments{All limits are given at 5-$\sigma$ significance.}

\tablenotetext{a}{Assumes a distance of 5~kpc to the pulsar and PWN.}

\tablenotetext{b}{Assumes that the pulsar 
was born $\Theta = 16\farcm1-20\farcm6$ east of its current location,
near the geometric center of SNR~\snr.}
\end{deluxetable}


\end{document}